\documentclass[aps,nofootinbib, 10pt,superscriptaddress]{revtex4-2}
\usepackage{graphicx}
\usepackage[T1,T2A]{fontenc}
\usepackage[utf8]{inputenc}
\usepackage[english]{babel}
\usepackage{amsmath}
\usepackage{amssymb}
\usepackage{perpage}
\MakePerPage{footnote}
\usepackage{caption}
\usepackage{comment}
\usepackage[lofdepth,lotdepth]{subfig}
\usepackage{float}
\usepackage{multirow}
\usepackage{stackengine}[2013-10-15]
\usepackage{dutchcal}
\usepackage[normalem]{ulem}
\usepackage{xcolor}

\addto\captionsenglish{}

\renewcommand{\Re}{\operatorname{Re}}
\renewcommand{\Im}{\operatorname{Im}}

\newcommand\spaceoverxrightarrow[2][\stackgap]{%
  \ensurestackMath{\stackon[#1]{\xrightarrow[\phantom{#2}]{}}%
  {\scriptscriptstyle #2}}}

\begin{document}

\preprint{INR-TH-2025-008}

\title{ CFT-approach to Rotating Field Lumps in Attractive Potential }

\author{Yulia Galushkina}
\affiliation{Institute for Nuclear Research of RAS, prospekt 60-letiya Oktyabrya 7a, Moscow, 117312}

\author{Eduard Kim}
\email{kim.e@phystech.edu}
\affiliation{Institute for Nuclear Research of RAS, prospekt 60-letiya Oktyabrya 7a, Moscow, 117312}
  \affiliation{Moscow Institute of Physics and Technology, Institutsky lane 9, Dolgoprudny, Moscow region, 141700}

\author{Emin Nugaev}
\affiliation{Institute for Nuclear Research of RAS, prospekt 60-letiya Oktyabrya 7a, Moscow, 117312}

\author{Yakov Shnir}
\affiliation{BLTP JINR, Joliot-Curie St 6, Dubna, Moscow region, 141980}

\begin{abstract}
We demonstrate the importance of relativistic corrections for the study of the stability of $(2+1)$-dimensional non-topological solitons with quartic self-interaction in the low-energy limit. This result is explained by the restoration of conformal symmetry in the non-relativistic limit. Particularly, the corresponding cubic Gross-Pitaevskii equation supports scale-free non-topological solitons. An unbroken conformal symmetry provides the additional degeneration that allows for the exact result for the energy and angular momentum of the stationary classical solutions. We study the violation of conformal symmetry by relativistic corrections.
The emergence of exponentially growing modes on the classical background is demonstrated using analytical approximations and numerical calculations.
  
\end{abstract}

\maketitle

\section{Introduction}
Non-topological solitons occur in complex scalar field theories possessing an unbroken continuous
global symmetry and a suitable interaction with additional fields \cite{Friedberg:1976me,Lee:1991ax}, see also recent reviews \cite{shnir2018topological,Nugaev:2019vru,Zhou:2024mea}. An important example is a simplified effective theory of a single complex self-interacting scalar field. In this case, the properties of non-topological solitons, which are usually referred to as Q-balls, can be studied in a more simple model \cite{Coleman:1985ki}.   
They are localized configurations with an explicitly time-dependent phase that are stabilized by the conserved Noether charge associated with this symmetry.
%The Q-balls have been discussed in
%a wide variety of possible  applications in particle physics, cosmology, and %astrophysics, for reviews, see, e.g., %\cite{Lee:1991ax,Radu:2008pp,shnir2018topological}. 
Physically, they may be viewed as a coherent state of the field quanta which correspond to an extremum of the effective
energy functional for a fixed value of the Noether charge. The energy of Q-ball depends in some way on the Noether charge $Q$, $E=m f(Q)$, where $m$ is the mass of the quanta of the scalar field. Thus, they are kinematically stable if $E<mQ$. However, there are also metastable and unstable solutions.  

%Certainly, the Q-balls resemble non-topological lumps in the atomic Bose-%Einstein condensate \cite{Enqvist:2003zb}

The Q-balls may be formed in a primordial phase transition contributing to various scenarios of the evolution of the early Universe \cite{Affleck:1984fy, Frieman:1989bx,Cotner:2019ykd}, and they are also considered as candidates for Dark Matter (DM) \cite{Kusenko:1997hj,Enqvist:1997si}. Coupling of scalar field to gravity gives rise to gravitating counterparts of Q-balls, so-called boson stars; see e.g., \cite{Jetzer:1991jr,Liebling:2012fv,Suarez:2013iw}. In this scope, numerical modeling shows that the stability issue of rotating lumps of the scalar field plays an important role in the description of ultralight DM, e.g. a rotating Bose star has additional instabilities \cite{Dmitriev:2021utv,Siemonsen:2020hcg}.

The Q-balls  appear to exist in arbitrary number of spatial dimensions $d$ \cite{Cervero:1986ec,Kumar:1987di,Battye:2000qj,Axenides:1999hs,Bowcock:2008dn,Bazeia:2015gkq,Alonso-Izquierdo:2023xni},
in $d\ge 2$ they admit spinning generalizations with an intrinsic angular
momentum \cite{Volkov:2002aj,Radu:2008pp}. 
The one- and two-dimensional Q-balls can be embedded in three dimensions, forming so-called 
Q-walls and  Q-tubes, respectively \cite{MacKenzie:2001av,Sakai:2010ny,Shnir:2011gr,Tamaki:2012yk,Nugaev:2014iva}. Recent analysis of $d=2$ dynamics of solitons \cite{Brax:2025uaw,Brax:2025vdh} shows the importance of taking self-action into account even for non-relativistic regimes. 

The symmetry analysis of field theory provides deep insights into properties of the fundamental interactions. Among a wide variety of physical symmetries, conformal symmetry stands out. It was established \cite{deKok:2007ur} that interacting non-relativistic field theory is invariant under space-time transformations of the Schrodinger group even in the presence of self-interaction term $|\psi|^{2k}$ if relation
\begin{equation*}
kd=d+2
\end{equation*}
is satisfied. In this case, common Galilei symmetry is complemented by scale-invariance and special conformal symmetry. Thus, in (1+1) dimensions self-interaction term $|\psi|^{6}$ that describes 3-body interaction gives rise to conformal scale-free bright solitons \cite{Galushkina:2025hkw}. In this paper, we consider planar solutions in $(2+1)$ dimensions that yield $|\psi|^{4}$ non-linearity, which corresponds to two-body interaction. In Sec.\ref{tube} we introduce our Lorentz-invariant model and provide analytical consideration of non-topological solitons with large winding numbers. Then, Sec.\ref{Stability} is devoted to the numerical analysis of the issue of these solutions.  We demonstrate that the decay rate of exponential instability vanishes in the non-relativistic limit. This result is explicitly discussed within the framework of non-relativistic conformal field theory in Sec.\ref{NR_sec}.

\section{Q-tube solutions}\label{tube}
Let us consider the (2+1)-dimensional Lorentz-invariant Lagrangian\footnote{We use natural system of units $\hbar=c=1$. Thus, the parameters of the theory are of units of mass $M$.} of a complex scalar field with quartic self-interaction
\begin{equation} 
\label{lagrangian}
    \mathcal{L} = \partial^{\mu}\phi^* \partial_{\mu}\phi- m^2 \phi^* \phi + \frac{\lambda}{2} (\phi^* \phi)^2,
\end{equation}
where \([m] = M\), \([\lambda] = M\), \(\lambda > 0\). This theory possess an unbroken global $U(1)$ symmetry and an attractive potential that may support soliton solutions of non-topological type. The solitons and their properties were firstly
studied in the $(3+1)$-dimensional model (\ref{lagrangian}) in the paper \cite{1970JMP....11.1336A}. 
To obtain non-topological solitons, we introduce the ansatz similar to one in this paper, but generalized for the rotating configurations:
\begin{equation}
\label{ansatz_background}
    \phi = e^{-{\rm{i}}\omega t} e^{{\rm{i}} n \theta} f(r),
\end{equation}
where \(r = \sqrt{\vec{x}^{2}}\) and $\theta$ are two-dimensional radial and angular coordinates, respectively, and \(n\) is a winding number. Now the corresponding field equation takes the form of an ordinary differential equation:
\begin{equation}\label{eq}
    f^{''}(r)+\frac{f^{'}(r)}{r} - \frac{n^2}{r^2} f(r) - (m^2 -\omega^2) f(r) + \lambda f^3(r) = 0.
\end{equation}
This equation has a symmetry \(\omega \to - \omega\), so for our numerical calculations we fix the sign of the frequency, \(\omega \geq 0\).  

At first hand, it is common that the stability of nontopological solitons may be examined by considering their \(U(1)\)-charge $Q$ and their energy $E$ \cite{Vakhitov:1973lcn,Friedberg:1976me}. Importantly, the conserving charge \(Q\) provides the non-topological stabilization of solitons. This value is given by the formula 
\begin{equation}
\label{Q}
Q = \int {\rm{i}}\left[ \phi^{*} \partial_0 \phi - \phi \partial_0 \phi^{*}\right] d^2x = 2 \pi \int^{\infty}_{0} 2\omega f^2(r) r dr.
\end{equation}
Furthermore, the energy of soliton \(E\) is provided by the integral
\begin{equation}
\label{E}
\begin{split}
&E = \int \left[\partial_0 \phi^*\partial_0 \phi + \partial_i \phi^*\partial_i \phi + m^2 \phi^*\phi - \frac{\lambda}{2} (\phi^*\phi)^2 \right]d^2x = \\
&\,\,\,\,\,\, 2 \pi \int^{\infty}_{0} \left[(\omega^2 + m^2 + \frac{n^2}{r^2}) f(r)^2 + (f'(r))^2 - \frac{\lambda}{2} f^4(r) \right] r dr.
\end{split}
\end{equation}
The spinning solutions also possess an intrinsic angular momentum $J$, which is proportional to the Noether charge, $J=n Q$ \cite{Volkov:2002aj,Radu:2008pp}.

For convenience, we introduce new dimensionless variables similarly as in \cite{1970JMP....11.1336A}
\begin{equation}
\label{scaling}
    \begin{split}
        & \tilde{r} = r \sqrt{m^2 - \omega^2} , \\
        & \tilde{f} = f\frac{\sqrt{\lambda}}{\sqrt{m^2 - \omega^2}}.
    \end{split}
\end{equation}
In terms of these variables, we obtain a field equation which does not depend on \(\omega\), \(\lambda\):

\begin{equation}\label{Scaled_eq}
    \tilde{f}^{''}(\tilde{r})+\frac{\tilde{f}^{'}(\tilde{r})}{\tilde{r}} - \left(1+\frac{n^2}{\tilde{r}^2} \right)\tilde{f}(\tilde{r}) + \tilde{f}^{3}(\tilde{r})= 0.
\end{equation}

In general, these equations can be solved only numerically, but for \(n \gg 1\) there might be a possibility of deriving an analytical approximation, as in the case of vortex solutions \cite{Bolognesi:2005zr}. For the purpose of finding an analytical solution, it is often useful to study integral condition
\begin{equation*}
    \int_{0}^{\infty}\left[ \tilde{f}^{''}\tilde{f}^{'}+\frac{\left(\tilde{f}^{'}\right)^2}{\tilde{r}} \right] d\tilde{r} = \int_{0}^{\infty}d\tilde{r}\left[\tilde{f}\tilde{f}^{'}\left(1+\frac{n^2}{\tilde{r}^2} \right)-\tilde{f}^{3}\tilde{f}^{'} \right],
\end{equation*}
which, along with known boundary conditions (see App.\ref{app.1}), results in the following relation
\begin{equation}\label{integral relation}
    \int_{0}^{\infty}d\tilde{r} \frac{\left(\tilde{f}^{'}(\tilde{r}) \right)^2}{\tilde{r}} = \int_{0}^{\infty}d\tilde{r}\frac{n^{2}\tilde{f}^{2}(\tilde{r})}{\tilde{r}^{3}}.
\end{equation}

Now we should recall another helpful tool: Eq.(\ref{Scaled_eq}) 
in a particle-mechanics analogy corresponds to Newton's dynamics with friction \cite{Coleman:1985ki,Volkov:2002aj}. A known asymptotic behavior of the Q-tube $\tilde{f}(\tilde{r})\sim \tilde{r}^{n}$ at $\tilde{r}\to 0$ means that in the limit of large $n$, the sharp peak of the soliton solution is drawn away from the origin. That fact and the exponential localization of the Q-tube mean that we can neglect the friction term in Eq.(\ref{Scaled_eq}) and the resulting equation is
\begin{equation}\label{limit eom}
    \tilde{f}^{''}(\tilde{r})=\tilde{f}(\tilde{r})\left(1+\frac{n^2}{R^2} \right) - \tilde{f}^{3}(\tilde{r}), \text{ for } n\to \infty,
\end{equation}
where $R$ is the radius of the solution's peak.

\begin{figure}[h]
\centering
\subfloat[Subfigure 1 list of figures text][]{
\includegraphics[width=0.45\textwidth]{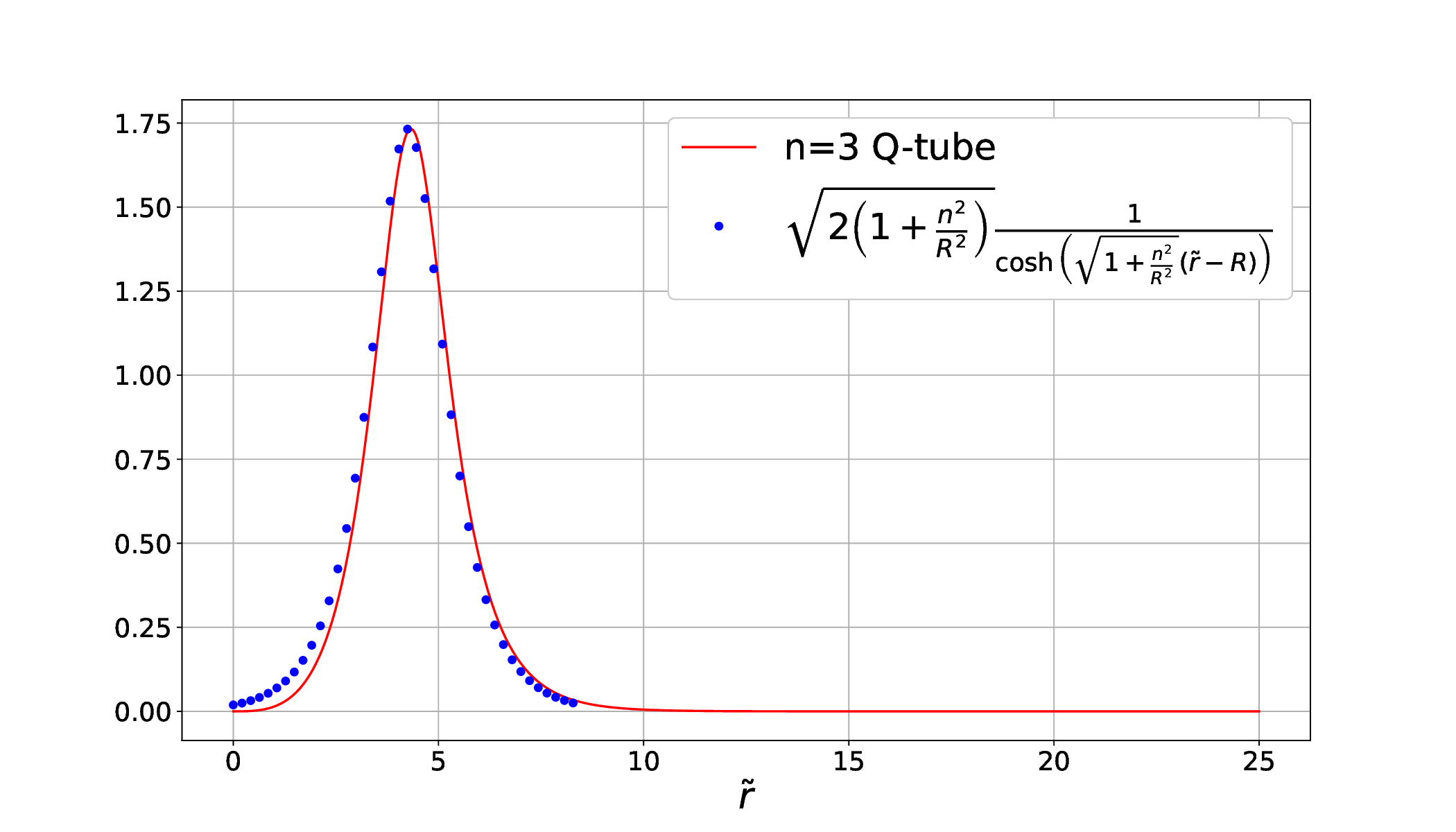}
\label{fig:subfig1}}
%\qquad
\subfloat[Subfigure 3 list of figures text][]{
\includegraphics[width=0.45\textwidth]{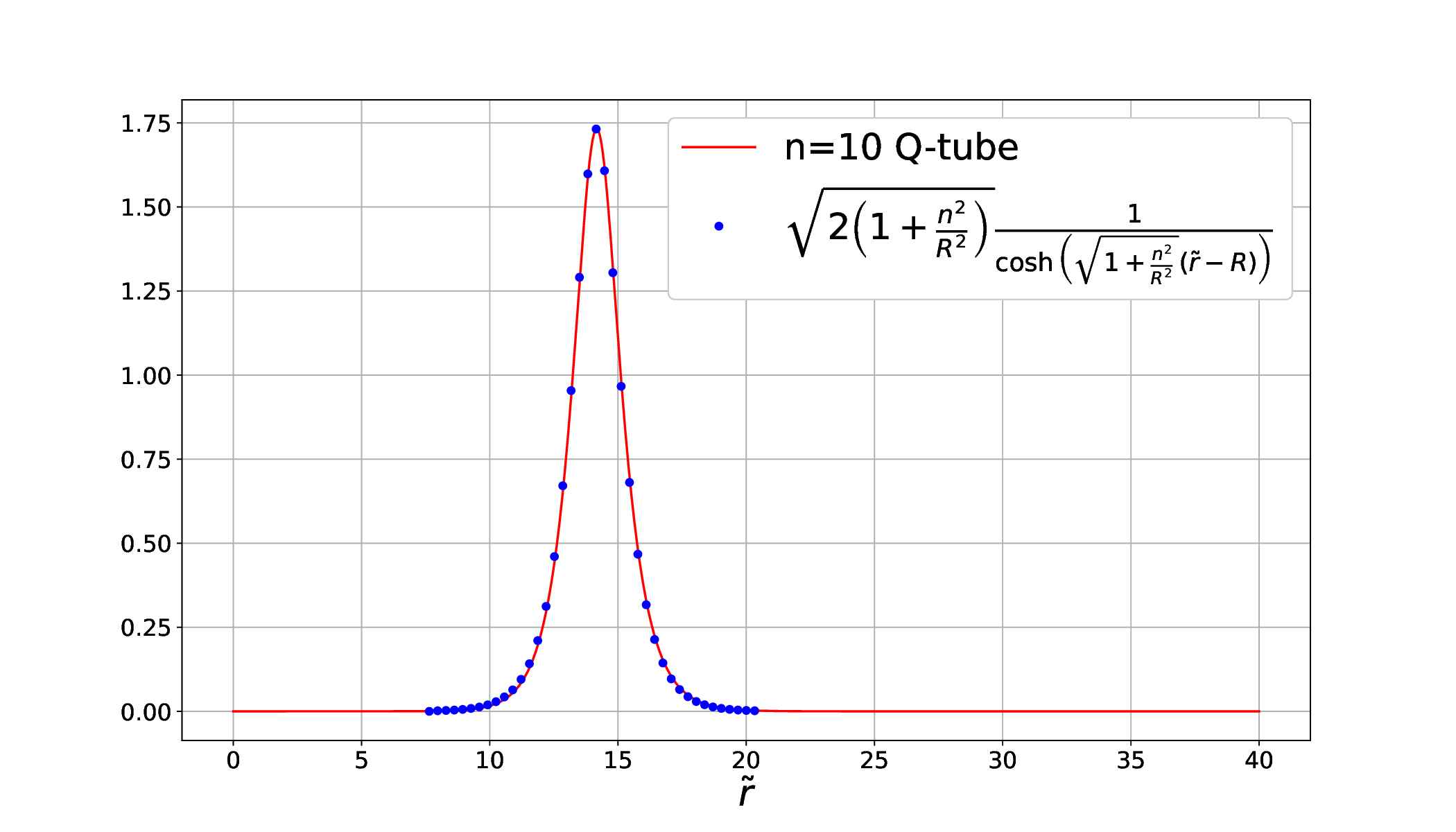}
\label{fig:subfig3}}
%\qquad
\caption{Comparison of numerical integration with an approximate analytical solution (\ref{analytical Q-tube}).}
\label{fig.1}
\end{figure}

Soliton solution of Eq.(\ref{limit eom}) has an exact analytical form
\begin{equation}\label{analytical Q-tube}
    \tilde{f}(\tilde{r}) = \sqrt{2\left(1+\frac{n^2}{R^2}\right)}\frac{1}{\cosh{\left(\sqrt{1+\frac{n^2}{R^2}}\cdot (\tilde{r}-R) \right)}}.
\end{equation}
In addition to that, using Eq.(\ref{integral relation}) we come to the conclusion that in the large $n$ limit parameter $R$ satisfies following relation
\begin{equation}
    \frac{n}{R} = \frac{1}{\sqrt{2}}.
\end{equation}
We have checked our numerical results by comparing them with our analytical solution (see Fig.\ref{fig.1}). 
%The presence of this approximation brings many conveniences besides gaining direct Q-tube profile, these include: ...

\begin{figure}[h!]
    %\centering
    \includegraphics[width=0.7\linewidth]{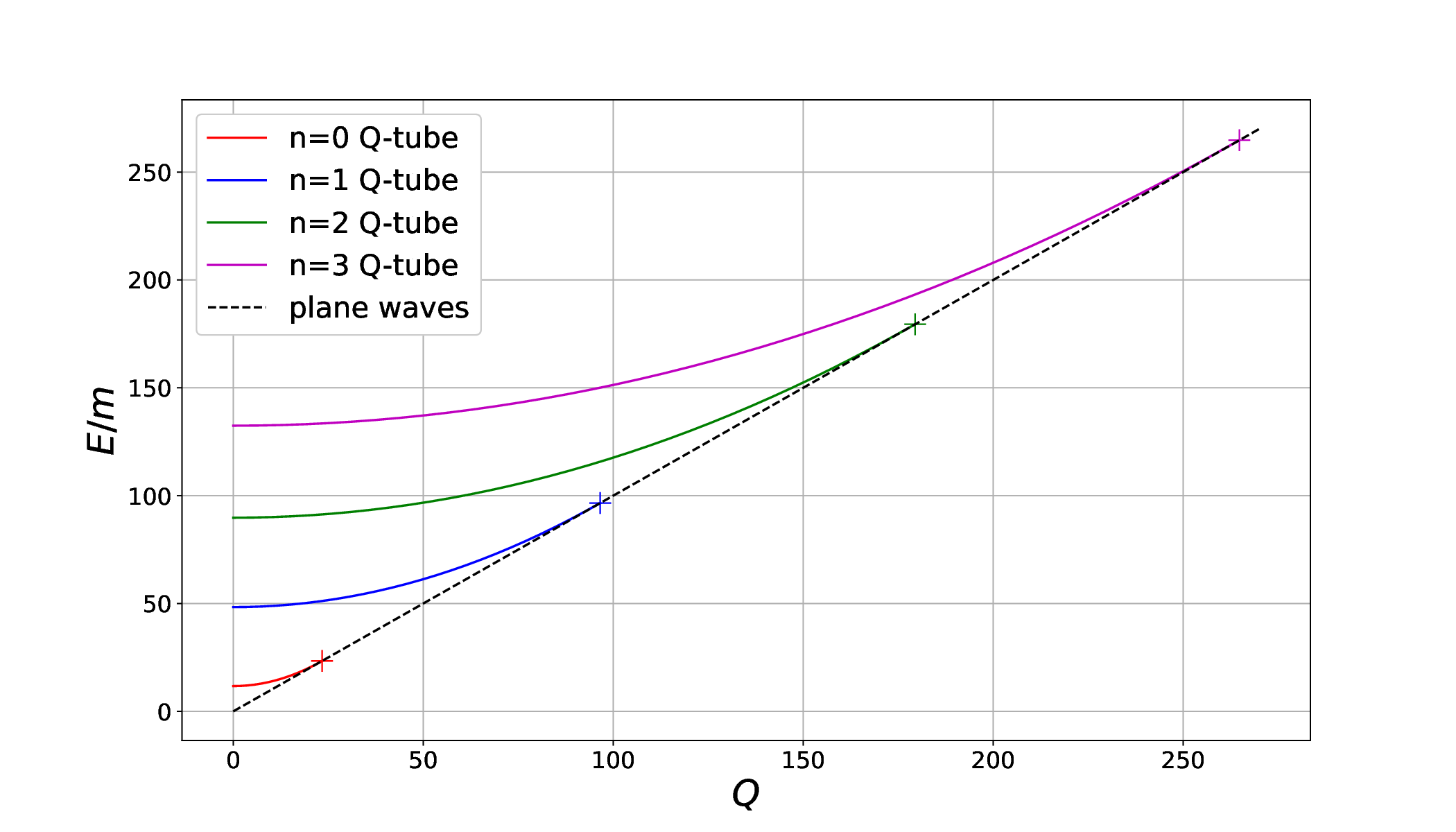}
    \caption{Integral characteristics of Q-tubes with different values of the parameter n. Calculations were performed by setting $\frac{\lambda}{m}=1$.}
    \label{fig.2}
\end{figure}

In terms of \(\tilde{f}\) and \(\tilde{r}\) we can obtain scaling formulas for \(Q\) and \(E\). Indeed, \(Q\) scales in the simplest way:
\begin{equation}\label{Scaled_Q}
    Q = \frac{\omega}{m} Q_{max},
\end{equation}
where
\begin{equation}\label{Q_max}
    Q_{max} = \frac{4 m \pi}{\lambda} \int^{\infty}_0 \tilde{f}^2(\tilde{r}) \tilde{r} d\tilde{r}
\end{equation}
is a finite dimensionless constant that can be evaluated numerically. One can notice that \(Q = Q_{max}\) at \(\omega = m\). The formula for the energy can be presented in the similar form, with the explicit dependence on \(\omega\):
\begin{equation}\label{Scaled_E}
    E = \frac{m^2 - \omega^2}{m^2}E_{0} + \frac{\omega^2}{m} Q_{max},
\end{equation}
where 
\begin{equation*}\label{E_{max}}
    E_{0} = \frac{2 m^2 \pi}{\lambda} \int^{\infty}_0 \left[ \left(1 + \frac{n^2}{\tilde{r}^2}\right) \tilde{f}^2(\tilde{r}) + (\tilde{f}'(\tilde{r}))^2 - \frac{1}{2} \tilde{f}^4(\tilde{r})\right] \tilde{r} d\tilde{r},
\end{equation*}
which is the energy of the static soliton (\(\omega = 0\)). The relation between values $E_0$ and $Q_{max}$ can be obtained using analytical results of App.\ref{app. symmetry}. Let us consider $d Q/ d\omega$ and $d E/d \omega$ using  (\ref{Scaled_Q}), (\ref{Scaled_E}). Then, the condition $d E/d \omega=\omega d Q/d \omega$ results in
\[
2m\omega Q_{max}-2\omega E_0-=m\omega Q_{max}.
\]
Thus, $E_0=m Q_{max}/2$.

One can see that the maximum value of \(E = mQ_{max}\) is achieved at \(\omega = m\). This fact, as well as the existence of \(Q_{max}\), will be explained in Sec.\ref{NR_sec} by considering non-relativistic limit of the theory. We will show that the restoration of conformal symmetry in non-relativistic limit determines these peculiar features of the model. 

Values of the parameters \(E_0\) and \(Q_{max}\) for these values of \(n\) are given in Table \ref{table}. The exact result $E_0=m Q_{max}/2$ is reproduced with high accuracy. 

The \(E(Q)\) dependence for \(n = 0, 1, 2 ,3\) is shown in Fig.\ref{fig.2}. One can see that according to the Vakhitov-Kolokolov criterion \cite{Vakhitov:1973lcn}, the solitons at \(\omega < m\) should be linearly unstable. Thus, we move on to studying their decay modes.

\begin{table}[h]
\centering
\caption{Critical integral characteristics of Q-tubes with different winding number $n$.}
\begin{tabular}{ |p{2cm}|p{4.5cm}|p{2cm}|  }
 \hline
 \multicolumn{3}{|c|}{Q-tubes with $\frac{\lambda}{m}=1$} \\
 \hline
 Parameter n & Energy of static solution $\frac{E_0}{m}$ & $Q_{max}$ \\
 \hline
 0 & 11.7009 & 23.4018 \\
 \hline
 1 & 48.29835 & 96.5967 \\
 \hline
 2 & 89.752 & 179.504 \\
 \hline
 3 & 132.4234 & 264.8468 \\
 \hline
\end{tabular}
\label{table}
\end{table}

\section{Q-tubes Stability}\label{Stability}

In this section, we study the spectrum of linear perturbations on the solitons described in the previous section. According to the Vakhitov-Kolokolov instability criterion 
\begin{equation}
    \frac{\omega}{Q}\frac{dQ}{d\omega}>0,
\end{equation}
it is justified to look for decay modes. Purely exponential decay modes are obtained using the following ansatz\footnote{For more general ansatz see \cite{1970JMP....11.1336A, Nugaev:2014iva}.}.
\begin{equation}
\delta\phi(t,\vec{x}) = e^{-i\omega t} e^{{\rm{i}}n\theta}e^{\gamma_{dec.}t}\left(\Re\xi(r)+i\Im\xi(r)\right),
\end{equation}
where \(\omega\), \(n\) are the parameters of background soliton.  

The equations for perturbations take the form
\begin{equation}\label{Lin. eqs decay modes}
    \begin{split}
        & \Re\xi'' + \frac{\Re\xi'}{\tilde{r}} -\frac{n^2}{\tilde{r}}\Re\xi = \frac{\left(m^{2}-\omega^{2}+\gamma_{dec.}^{2}\right)\Re\xi+2\omega\gamma_{dec.}\Im\xi}{m^{2}-\omega^{2}} - 3\tilde{f}^{2}\Re\xi , \\
        & \Im\xi'' + \frac{\Im\xi'}{\tilde{r}} -\frac{n^2}{\tilde{r}}\Im\xi = \frac{\left(m^{2}-\omega^{2}+\gamma_{dec.}^{2}\right)\Im\xi-2\omega\gamma_{dec.}\Re\xi}{m^{2}-\omega^{2}} - \tilde{f}^{2}\Im\xi .
    \end{split}
\end{equation}

\begin{figure}[htb!]
    \centering
    \includegraphics[width=0.7\linewidth]{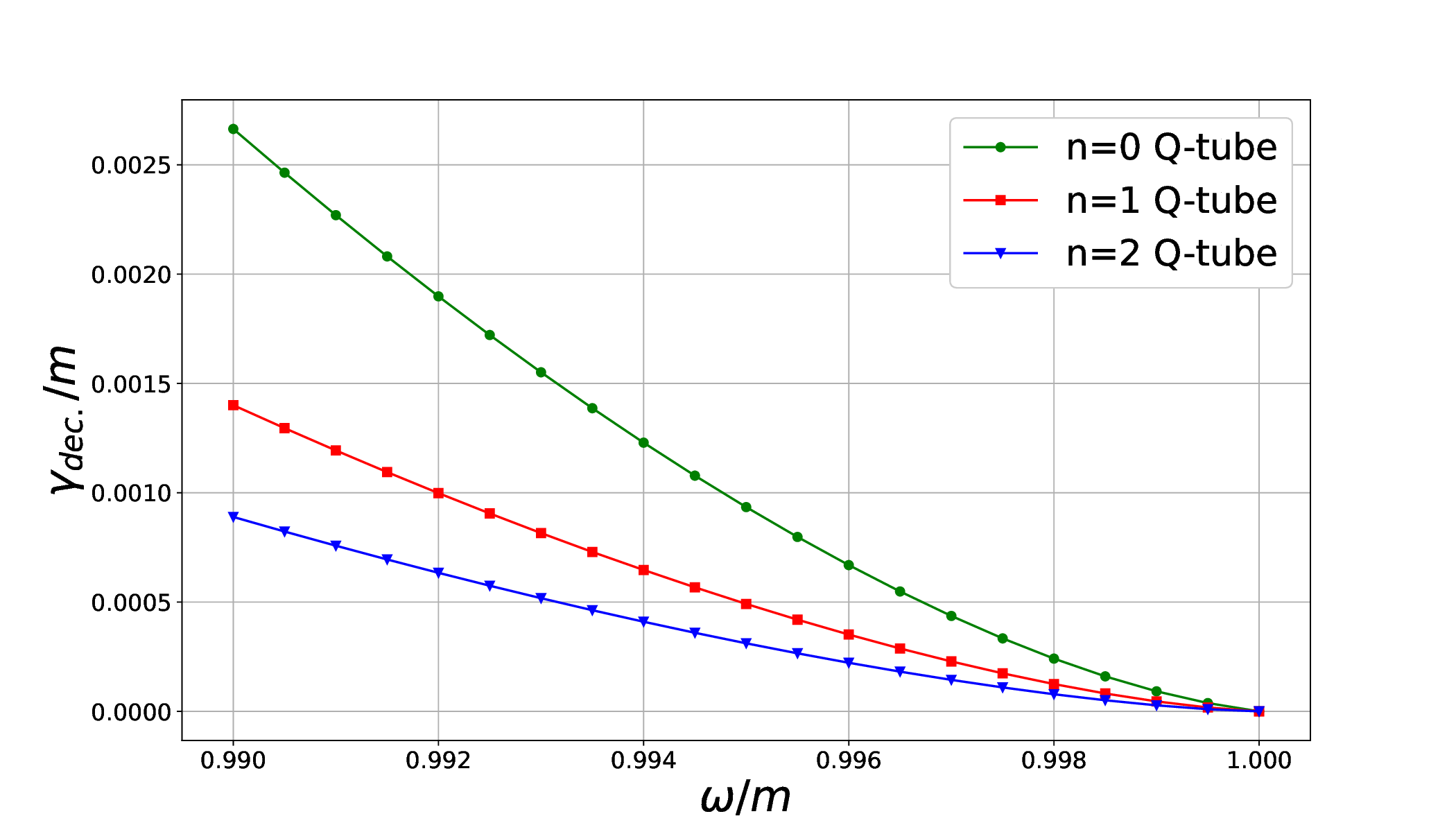}
    \caption{Decay parameter \(\gamma_{dec.}/m\) for \(\omega/m \in [0.99, 1]\), \(n = 0,1,2\).}
    \label{modes}
\end{figure}

\begin{figure}[htb!]
    \centering
    \includegraphics[width=0.7\linewidth]{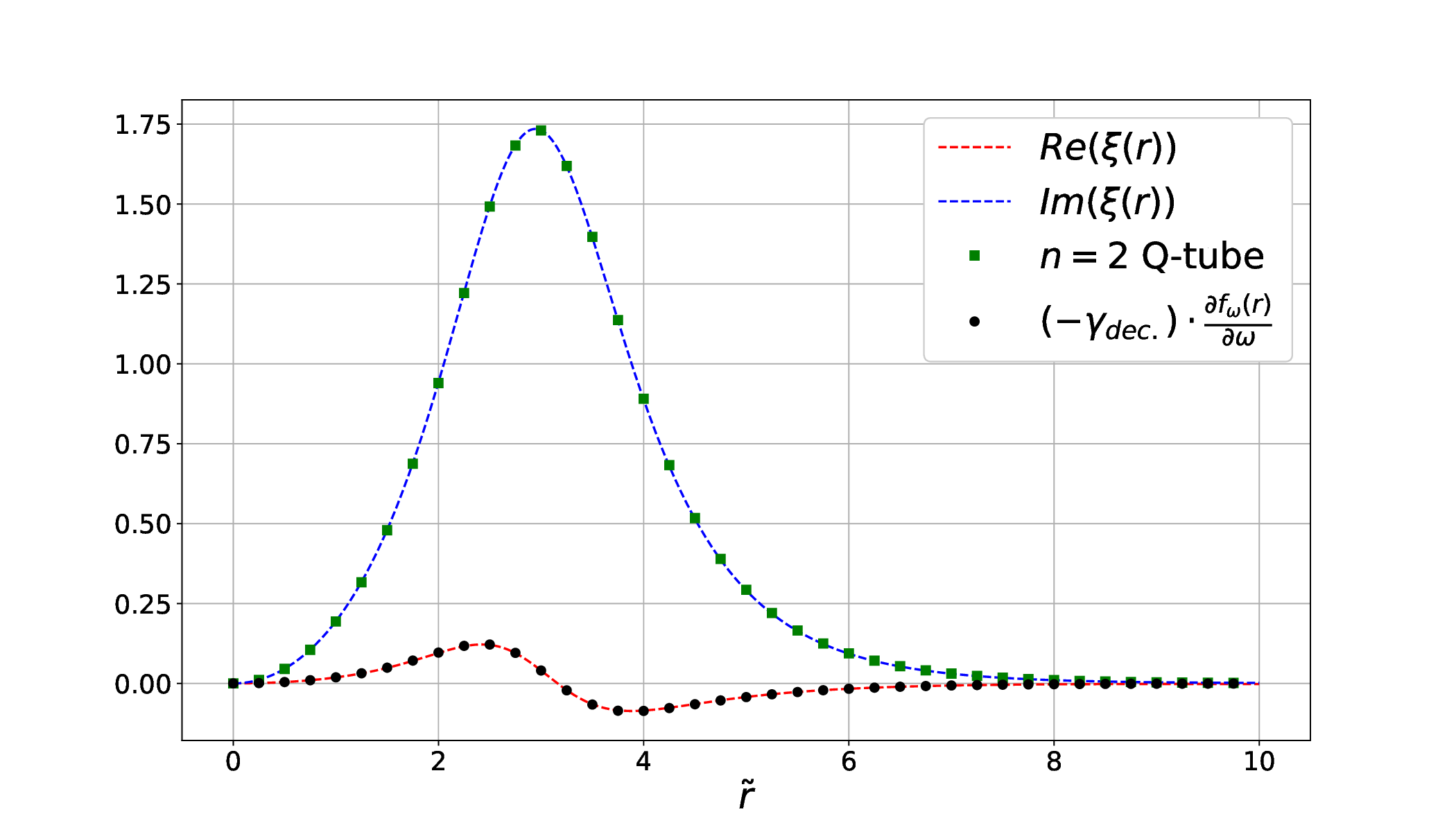}
    \caption{Decay mode profile at $\omega/m=0.995$ and $\gamma_{dec.}/m=3.11\cdot10^{-4}$. Scaled soliton profile and $\displaystyle{(-\gamma_{dec.})\frac{\partial f_{\omega}(r)}{\partial \omega}}$ are added for comparison.}
    \label{decay mode}
\end{figure}

Numerical results for the decay modes spectrum are shown in Fig.\ref{modes}. Importantly, in the limit $\omega\to m$ parameter $\gamma_{dec.}$ vanishes. When $\frac{\gamma_{dec.}}{\omega}\ll 1$ decay modes might be approximated by expanding a background solution in perturbation series as
\begin{equation}\label{Perturbation series}
    i\delta \phi(t,\vec{x})=ie^{-i\left(1+i\frac{\gamma}{\omega} \right) \omega t}e^{{\rm{i}}n\theta}f_{1+i\frac{\gamma}{\omega}}(r)\approx e^{-i\omega t}e^{{\rm{i}}n\theta} \left( 1 + \gamma t \right)\left(if_{\omega}(r) -\gamma\frac{\partial f_{\omega}(r)}{\partial \omega}  \right),
\end{equation}
where we use the notation \(f_{\omega}\) to stress that the solution \(f\) of Eq.(\ref{eq}) parametrically depends on \(\omega\).
Despite the absence of a cusp \(dQ/d\omega = 0\), the validity of Eq.(\ref{Perturbation series}) is justified by the numerical simulation for different values of \(n\) (e.g. see Fig.\ref{decay mode} for \(n = 2\)). The reason why this expansion is valid will be formulated in Sec.\ref{NR_sec} while considering the non-relativistic limit of the theory. Moreover, we will show that the behavior of \(\gamma_{dec.}(\omega)\) in the limit \(\omega \to m\) is determined by the restored conformal symmetry.

Note that for the larger \(n\) the soliton is more stable (the decay parameter \(\gamma_{dec.}\) decreases with larger $n$), although the energy of soliton \(E\) is larger (see Figs.\ref{fig.2}-\ref{modes}). Thus, the higher angular momentum reduces the instability of the configurations in the attractive potential.

\section{Non-relativistic Limit and Conformal Symmetry}
\label{NR_sec}

The results of the previous section become clearer while considering the non-relativistic limit of the theory (\ref{lagrangian}). For this purpose we use substitution 
\begin{equation}\label{NR ansatz}
    \phi(t,\vec{x}) = \frac{1}{\sqrt{2m}}e^{-imt}\psi(t,\vec{x}).
\end{equation}
It can be seen that the substitution (\ref{NR ansatz}) leads to the vanishing of the mass term in the Lagrangian
\begin{equation}\label{NR sub. L}
    \mathcal{L} = \frac{1}{2m}\left[ |\dot{\psi}|^{2} + im(\psi^{\ast}\dot{\psi} - \psi\dot{\psi}^{\ast}) \right] - \frac{1}{2m}\nabla\psi^{\ast}\nabla\psi + \frac{\lambda}{8m^{2}}(\psi^{\ast}\psi)^{2}.
\end{equation}
\begin{comment}Then, the kinetic term of the Lagrangian (\ref{lagrangian}) contains two terms, $|\dot\psi|^2/m$ and $i\psi^*\dot\psi$. Indeed, additional term without time derivatives is canceled by the mass term. For solutions with $\dot\psi\ll m\psi$ we can omit $|\dot\psi|^2/m$ and write down the non-relativistic Lagrangian
\end{comment}
Using relativistic ansatz (\ref{ansatz_background}) that describes stationary configurations one can check that the time dependence of $\psi(t,\vec{x})$ is $\psi(t,\vec{x})\propto e^{-i(\omega-m)t}$. This allows to compare terms in square brackets in Lagrangian (\ref{NR sub. L}): $|\dot{\psi}|^{2}=|(m-\omega)^{2}\psi^{2}|$ and $|m\psi^{\ast}\dot{\psi}| = |m(m-\omega)\psi^{2}|$. In the limit $\omega \to m$, the ratio $\displaystyle{\frac{|\dot\psi|^2}{|m\psi^{\ast}\dot{\psi}|}}$ is suppressed as $\frac{m-\omega}{m}$. Thus, the non-relativistic Lagrangian is written as
\begin{equation}\label{NR lagrangian}
    \mathcal{L}_{NR} = i\psi^{\ast}\dot{\psi} - \frac{1}{2m}\nabla\psi^{\ast}\nabla\psi + \frac{\lambda}{8m^{2}}\left(\psi^{\ast}\psi \right)^{2},
\end{equation}
which results in the nonlinear Schr\"{o}dinger equation. In this scope, dimensionless combination $\displaystyle{\frac{m-\omega}{m}}$ controls relativistic corrections for stationary configurations. Discussion of a more general approach to construction of non-relativistic limit was provided in \cite{Namjoo:2017nia}.
%\ek{In the limit $\omega\to m$, the binding energy per constituent $d(mQ-E)/dQ=(m-\omega)$ tends to zero, which represents a system of particles with mass $m$ near the rest state (i.e. non-relativistic regime).}
%It can be seen, that soliton solutions in theory (\ref{NR lagrangian}) correspond to relativistic Q-tubes of Eq.(\ref{eq}) in the limit $\omega \to m$.

%One can see that the solutions of this equation 

%\(\omega \to m\) are reproduced by the solutions of the non-linear Schr\"{o}dinger equation.}

\begin{figure}[htb]
    \centering
    \includegraphics[width=0.7\linewidth]{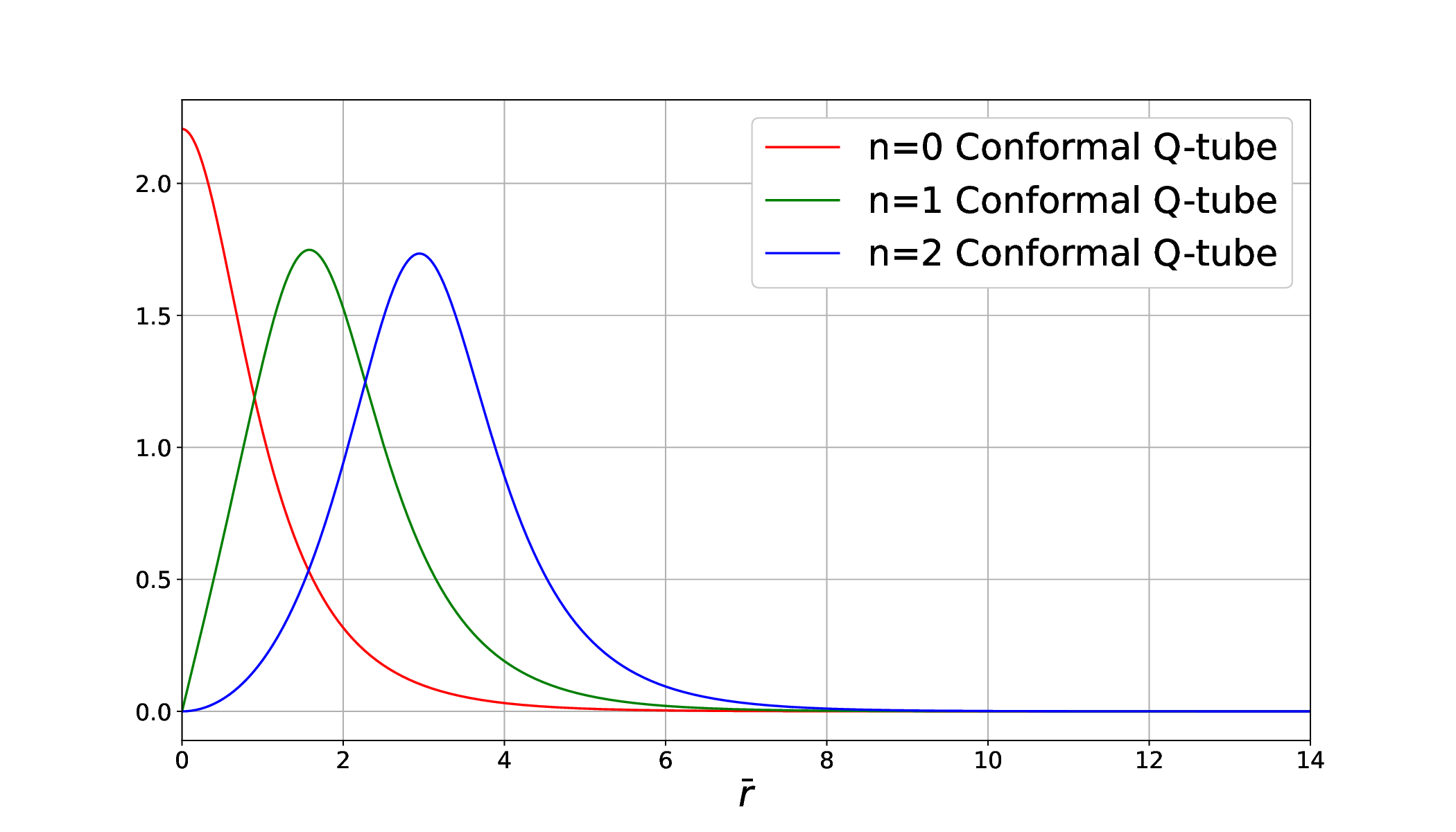}
    \caption{Conformal Q-tube profiles $\displaystyle{\frac{\bar{h}(\bar{r})}{\sqrt{2m}}}$ with winding number $n=0,1,2$.}
    \label{Conformal Q-tubes}
\end{figure}

In (2+1) dimensions, the theory (\ref{NR lagrangian}) is invariant under the Schr\"{o}dinger group. In particular, we are interested in the scale-invariance (dilatations) and the special conformal symmetry that are exact and unbroken in the presence of the quartic self-interaction term \cite{deKok:2007ur}. The corresponding equations of motion support conformal Q-tube solutions of the form $\psi(t,\vec{x}) = e^{i\mu t}e^{in\theta}h(r)$, where $\mu=m-\omega$. The equation of motion is written as
\begin{equation}
    h^{''}(r)+ \frac{h^{'}(r)}{r}-\frac{n^2}{r^2}h(r)=2m\mu h(r)-\frac{\lambda}{2m}h^{3}(r).
\end{equation}
This equation allows for scaling
\begin{equation}
    \begin{split}
        & \bar{r} =r\sqrt{2m\mu}~ , \\
        & \bar{h} = h\sqrt{\frac{\lambda}{2m\mu}}
    \end{split}
\end{equation}
and can be rewritten into
\begin{equation}
    \bar{h}^{''}(\bar{r})+ \frac{\bar{h}^{'}(\bar{r})}{\bar{r}}-\frac{n^2}{\bar{r}^2}\bar{h}(r)= \bar{h}(\bar{r})-\frac{1}{2m}\bar{h}^{3}(\bar{r}).
\end{equation}
One can solve this equation numerically and find conformal Q-tube solutions (see Fig.\ref{Conformal Q-tubes}).

By analogy with similar solutions constructed in 
\cite{Galushkina:2025hkw}, these Q-tubes are scale-free and possess a constant global $U(1)$ charge $N=const$. In our case, $N=Q_{max}$, where $Q_{max}$ can be obtained from Table \ref{table}. The Hamiltonian of the model (\ref{NR lagrangian})
\begin{equation}
    H = 2\pi \int_{0}^{\infty} \left[\frac{|\nabla\psi|^{2}}{2m}-\frac{\lambda}{8m^{2}}|\psi|^{4} \right]rdr
\end{equation}
corresponds to the planar Gross-Pitaevskii theory. Remarkably, $H=0$ for soliton solutions with an arbitrary $n$. One can see that the latter is a result of an unbroken scale invariance and conformal symmetry. Indeed, the corresponding symmetry generators $D$ and $K$ are expressed as follows
\begin{equation*}
\begin{split}
    & D = 2tH + \frac{i}{2}\int \vec{x}\left(\psi^{\ast}\vec{\nabla}\psi - \psi\vec{\nabla}\psi^{\ast} \right)d^{2}x, \\
    & K = t^{2}H - tD - \frac{m}{2}\int \vec{x}^{2}(\psi^{\ast}\psi) d^{2}x,
\end{split}
\end{equation*}
and for stationary configurations we obtain
\cite{deKok:2007ur}
\begin{equation}
    \frac{dK}{dt}=-t\frac{dD}{dt},\qquad \frac{dD}{dt}=2H.
\end{equation}
From the other side, these generators are conserved due to the symmetries of the model. Thus, $H=0$ is the only possible realization for static/stationary solutions of theory (\ref{NR lagrangian}) in $(2+1)$ dimensions (for more details see App.\ref{app. symmetry}). In other words, the properties of Q-tube solutions are independent of the dimensionful parameter $\mu$ due to unbroken dilatations and conformal symmetry. This analytical result is consistent with the numerical zeroing out of $(E-mQ)$ (see Fig.\ref{fig.2}) in the critical point $\omega=m$, which was obtained in section \ref{tube}.

\begin{figure}
    \centering
    \includegraphics[width=0.7\linewidth]{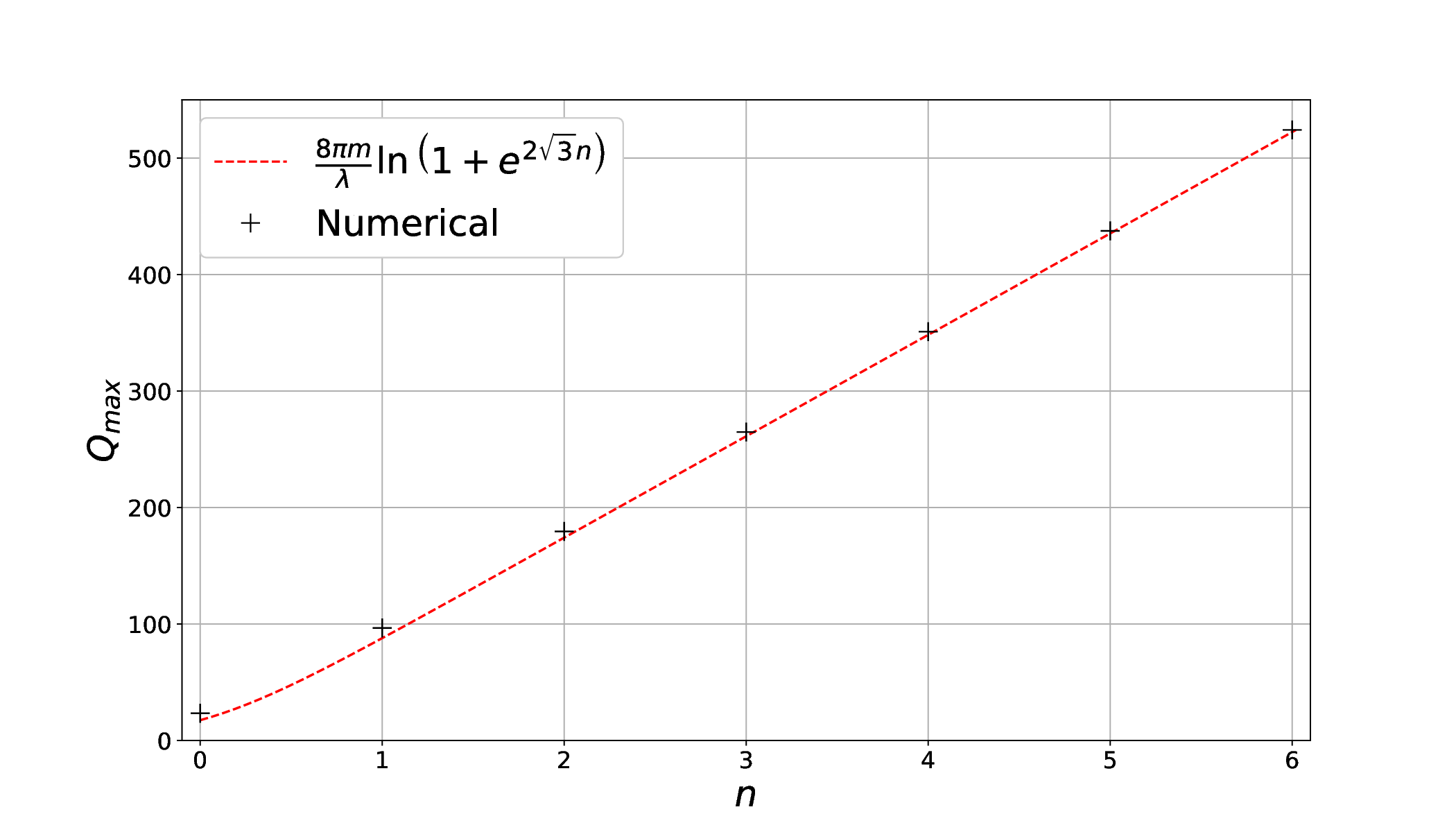}
    \caption{The value of $Q_{max}$ plotted versus winding number $n$ at $\lambda/m=1$. Cross markers indicate results of numerical integration, while analytical estimations are represented by a dashed line.}
    \label{Qmax Comparison}
\end{figure}

It was noted that the non-relativistic global $U(1)$ charge 
\begin{equation}
    N = \int_{-\infty}^{\infty} |\psi(t,\vec{x})|^{2} dx^2 =  2\pi \int_{0}^{\infty} \bar{h}(\bar{r})^{2} \bar{r}d\bar{r} = Q_{max}.
\end{equation}
where $Q_{max}$ (\ref{Q_max}) could be estimated using the analytical approximation (more details in App.\ref{app.1}). Thus, one can derive asymptotical formula 
%In terms of $Q_{max}$ (\ref{Q_max}) using the solution (\ref{analytical Q-tube}), one can derive asymptotical formula

\begin{equation}\label{Q_max analytical}
Q_{max}\approx\frac{8\pi m}{\lambda} \ln{\left(1+e^{2\sqrt{3}n} \right)} \spaceoverxrightarrow{n\to\infty} \frac{16\sqrt{3}\pi m}{\lambda}n.
\end{equation}
A direct comparison with numerical integration (see Fig.\ref{Qmax Comparison}) has shown that the accuracy of the analytical approximation (\ref{Q_max analytical}) exceeds $99\%$ at $n=4$ and continues to grow.
Remarkably, the angular momentum $J$ is proportional to $n^2$ (see App.\ref{app. symmetry}) and this result is valid for non-relativistic configurations even for the theories with slightly broken conformal invariance. 

%This result is in agreement with the following reasoning: the only natural parameter that can exhibit the plausibility of the solution (\ref{analytical Q-tube}) is $\displaystyle{1/ n^{2}}$.   

Relativistic solitons reproduce conformal solutions in the limit $\omega\to m$. Numerical results confirm the non-relativistic expansion on the parameter $\frac{m-\omega}{m}$. Moreover, the results of our numerical scanning show that the spectrum of linear perturbations on conformal Q-tubes contains only zero modes. In order to check this, we have implied numerical scanning (we introduced our method in \cite{Galushkina:2024iad}) of modes with known boundary conditions by using the decay modes ansatz
\begin{equation}
    \delta \psi(t,\vec{x}) = e^{i\mu t+\gamma_{dec.}t} e^{in\theta}\left(\Re{\xi} + i\Im{\xi} \right) 
\end{equation}
and linearized equations of motion
\begin{equation}\label{lin. nr eom}
    \begin{split}
        & \Re\xi'' + \frac{\Re\xi'}{\tilde{r}} -\frac{n^2}{\tilde{r}}\Re\xi = \Re\xi + \frac{\gamma_{dec.}}{\mu}\Im\xi - \frac{3}{2m}\bar{h}^{2}\Re\xi , \\
        & \Im\xi'' + \frac{\Im\xi'}{\tilde{r}} -\frac{n^2}{\tilde{r}}\Im\xi = \Im\xi - \frac{\gamma_{dec.}}{\mu}\Re\xi - \frac{1}{2m}\bar{h}^{2}\Im\xi .
    \end{split}
\end{equation}

The existence of a trivial spectrum is in accordance with Vakhitov-Kolokolov criterion
\begin{equation}\label{VK N.R.}
   \frac{\mu}{N}\frac{dN}{d\mu}=0.
\end{equation}
Indeed, our numerical scanning failed to find any decay solutions of linearized Eqs.(\ref{lin. nr eom}) apart from symmetry-related zero modes in theory (\ref{NR lagrangian}). 
%this relation indicates that the presence of conformal symmetry forbids any other modes but symmetry-related ones. 
Weak violation of conformal symmetry by relativistic generalization allowed us to generate decay modes near the isolated point as

\begin{equation}\label{Perturbation series nr}
    i\delta \phi(t,\vec{x})\approx e^{-i\omega t}e^{in\theta} \left( 1 + \gamma t \right) \left(if_{\omega}(r) -\gamma\frac{\partial f_{\omega}(r)}{\partial \omega}  \right)=e^{-i\omega t} e^{{\rm{i}}n\theta}(1+\gamma_{dec.}t)\left(\Re\xi(\tilde{r})+i\Im\xi(\tilde{r})\right).
\end{equation}
where we use the notation \(f_{\omega}\) to stress that the solution \(f\) of Eq.(\ref{eq}) parametrically depends on \(\omega\).
In the limit $\omega \to m$, parameter $\gamma_{dec.}$ tends to zero following Eq.(\ref{VK N.R.}) and this mode turns into the $U(1)$ symmetry zero mode. Although weak violation of conformal symmetry results in decay modes with small $\gamma_{dec.}/m$ this leads to drastically different dynamical behavior of solitons which become exponentially unstable. This type of instability is forbidden in purely non-relativistic conformal theory. Thus, we have demonstrated the importance of relativistic corrections for studying stability of low-energy Q-tubes in our model.  

\section{Outlook}
% Time in Outlook
In this paper, we have studied properties of Q-tube solutions with different values of the winding number \(n\) in a theory of complex scalar field with quartic self-interaction. For large \(n\), the solitons may be approximated by an analytical solution similar to one in (1 + 1)-dimensional theory.

Following a semi-analytical approach based on the scaling (\ref{Scaled_Q}, \ref{Scaled_E}), we considered integral characteristics of these solutions, such as energy functional and global $U(1)$ charge. Interestingly, at $\omega= m$, both quantities reach their finite maximal values which increase for larger \(n\). At $\omega=m$ the Q-tube's properties were interpreted within the framework of non-relativistic field theory. In this limit, the conformal symmetry is restored. 

The analysis of integral characteristics demonstrated that our solitons are kinematically unstable. To examine their instability, we studied their linear decay modes. The parameter of exponential decay \(\gamma_{dec.}\) decreases with larger \(\omega\) and vanishes at \(\omega = m\). The solitons with larger \(n\) are more stable (in terms of classical linear stability, described by the decay rate \(\gamma_{dec}\)) despite having higher energy. The orbital moment has a stabilizing effect on the solitons.

For better understanding of the features of solitons at $\omega \to m$, we studied the non-relativistic limit of our theory. We found that the behavior of the theory in this limit is very uncommon due to the restoration of the conformal symmetry. Firstly, the conformal theory describes scale-invariant branch of solitons (for a given value of \(n\)) that reproduces properties of the relativistic solution at the isolated point \(\omega = m\). All the solitons of the branch have the same charge and their non-relativistic Hamiltonian is identically equal to zero. Secondly, due to the conformal symmetry, the spectrum of linear perturbations is trivial and contains only zero modes. Thus, taking into account relativistic corrections is crucial for the study of the stability issue in this peculiar model.

One can apply our method of study of classical solutions in the theory with slightly broken conformal invariance for different problems of the field theory. Indeed, Lagrangian (\ref{NR lagrangian}) corresponds to the planar Gross-Pitaevskii theory. In addition to relativistic corrections, there are other mechanisms of conformal symmetry breaking. First of all, one can consider $d=3$ theory and study instabilities along the additional direction for the problems of evolution of rotating clumps of DM \cite{Brax:2025uaw}. One can also study D-term \cite{Mai:2012yc}
for solitons in conformal model to examine properties of the energy-momentum tensor.

Another possibility to slightly violate the conformal symmetry
is to add some interaction with appropriate field for $d=2$ theory. For example, it may be the interaction with a gauge field as in the model \cite{Jackiw:1990tz} for the special choice of the coupling constant. Additionally, it is an interesting question whether the conformal symmetry may remain unbroken after accounting for quantum corrections, as in \cite{deKok:2008ge}. 

%We demonstrated that a simple non-relativistic consideration of $(2+1)$-dimensional non-topological solitons with quartic self-interaction is insufficient for the study of stability issue. Conformal symmetry violation in the relativistic generalization of our model results in the emergence of exponentially growing modes on the classical background. 

\section*{Acknowledgments}
We thank Dmitry Levkov, Andrey Shkerin and Mikhail Smolyakov for valuable discussions.
Numerical studies of relativistic decay modes were supported by the grant RSF 22-12-00215-$\Pi$. The work of E. Kim on analytical approximations was supported by the Foundation for the Advancement of Theoretical Physics and Mathematics BASIS.

%\section*{Data Availability}
%The data that support the findings of this article are openly available \cite{galushkina2025cftapproachrotatingfieldlumps}.

\appendix
%\begin{comment}

\section{Large $n$ approximation}\label{app.1}

In this appendix, we consider an analytical form of Q-tubes in the limit of a large winding number $n$. Let us remind that the scaled equation of motion (\ref{limit eom}) is written as
\begin{equation}\label{A1}
    \tilde{f}^{''}(\tilde{r})+\frac{\tilde{f}^{'}(\tilde{r})}{\tilde{r}}=\tilde{f}(\tilde{r})\left(1+\frac{n^2}{\tilde{r}^2} \right) - \tilde{f}^{3}(\tilde{r}).
\end{equation}
For the purpose of finding an analytical solution, it is often useful to study the first integral
\begin{equation}
    \int_{0}^{\infty}\left[ \tilde{f}^{''}\tilde{f}^{'}+\frac{\left(\tilde{f}^{'}\right)^2}{\tilde{r}} \right] d\tilde{r} = \int_{0}^{\infty}d\tilde{r}\left[\tilde{f}\tilde{f}^{'}\left(1+\frac{n^2}{\tilde{r}^2} \right)-\tilde{f}^{3}\tilde{f}^{'} \right],
\end{equation}
which, along with known boundary conditions
\begin{equation*}
    \tilde{f}(\tilde{r})|_{\tilde{r}=0}=0 \quad \text{and} \quad \tilde{f}(\tilde{r})|_{\tilde{r}=\infty}=0,
\end{equation*}
results in
\begin{equation}\label{integral relation app.}
    \int_{0}^{\infty}d\tilde{r} \frac{\left(\tilde{f}^{'}(\tilde{r}) \right)^2}{\tilde{r}} = \int_{0}^{\infty}d\tilde{r}\frac{n^{2}\tilde{f}^{2}(\tilde{r})}{\tilde{r}^{3}}.
\end{equation}

Now we should recall another helpful tool: a mechanical interpretation. A known asymptotic behavior of Q-tube $\tilde{f}(\tilde{r})\sim \tilde{r}^{n}$ at $\tilde{r}\to 0$ means that in the limit of large $n$, the center of the soliton solution is drawn away from the origin. That fact and the exponential localization of Q-tube means that we can neglect the "friction" term in Eq.\ref{A1} and the resulting equation is
\begin{equation}
    \tilde{f}^{''}(\tilde{r})=\tilde{f}(\tilde{r})\left(1+\frac{n^2}{R^2} \right) - \tilde{f}^{3}(\tilde{r}), \text{ for } n\to \infty,
\end{equation}
where $R$ is the center of the solution. This equation has an exact solution in the form of a non-topological soliton
\begin{equation}\label{analytical Q-tube app.}
    \tilde{f}(\tilde{r}) = \sqrt{2\left(1+\frac{n^2}{R^2}\right)}\frac{1}{\cosh{\left(\sqrt{1+\frac{n^2}{R^2}}\cdot (\tilde{r}-R) \right)}}.
\end{equation}
Moreover, solution (\ref{analytical Q-tube app.}) can be put in Eq.(\ref{integral relation app.}) (along with considering $\tilde{r}=R$ in denominators) and we come to the result that for large $n$ the following relation holds
\begin{equation}
    \frac{n}{R} = \frac{1}{\sqrt{2}}.
\end{equation}
%We have checked our numerical results by comparing them with our analytical solution (see Fig.\ref{fig.1}). 
We neglect the friction term $\tilde f^{'}/\tilde{r}$
\begin{equation*}
    -\frac{\sqrt{2}}{\tilde{r}} \left(1+\frac{n^2}{R^2}\right) \tanh \left(\sqrt{1+\frac{n^2}{R^2}} \cdot(\tilde{r}-R)\right) \text{sech}\left(\sqrt{1+\frac{n^2}{R^2}}\cdot (\tilde{r}-R)\right)
\end{equation*}
because in the limit of the large winding number $n$ the soliton is localized around $\tilde{r}=R$ and is drawn away from the origin. Thus, essential information about the Q-tube is revealed by considering large $R$ limit. While the term $n^2/\tilde{r}^2$ remains relevant because large $\tilde{r}$ is compensated by large $n$, the friction term is suppressed with growing $\tilde{r}$. Thus, we have justified our approximation in the limit of large $n$. 

Analytical approximation (\ref{analytical Q-tube app.}) can be put directly into Eq.(\ref{Q_max}) for estimations of $Q_{max}$ value in the limit of large $n$
\begin{equation}
    Q_{max}\approx \frac{12m\pi}{\lambda}\int_{0}^{\infty}\frac{1}{\cosh^{2}{\left(\sqrt{\frac{3}{2}}\cdot(\tilde{r}-\sqrt{2}n) \right)}}\tilde{r}d\tilde{r} = \frac{8\pi m}{\lambda}\ln{\left(1+e^{2\sqrt{3}n}\right)}.
\end{equation}

\section{Symmetry-related exact results}
\label{app. symmetry}

In this appendix, we provide a discussion of some exact results for symmetries of our models. Firstly, we argue that the presence of an unbroken scale invariance in theory (\ref{NR lagrangian}) secures that any static/stationary solution of equations of motion has a zero-valued Hamiltonian. The scale invariance is a continuous symmetry, hence it corresponds to the conserving in time Noether charge. Scale invariance corresponds to space-time transformations
\begin{equation*}
    \vec{x}\to \sigma \vec{x},\quad t\to \sigma^{2} t,
\end{equation*}
which are complemented by the field transformation $\psi\to \sigma^{-1}\psi$. From the variation of the action $S=\int dtd^{2}x \mathcal{L}$, that is $\delta S = 0$ we can extract current conservation
\begin{equation*}
    \partial_{\mu}\left( \frac{\partial \mathcal{L}}{\partial(\partial_{\mu}\psi^{\ast})}\delta\psi^{\ast} + \frac{\partial \mathcal{L}}{\partial(\partial_{\mu}\psi)}\delta\psi - \mathcal{L}\delta x^{\mu} \right) = \partial_{\mu}J^{\mu} = 0,
\end{equation*}
where $x^{\mu}=(t,\vec{x})$. Space integration of this equation results in the following conservation law for Noether charge
\begin{equation*}
    \int d^{2}x \left(\frac{\partial }{\partial t}J^{0} + \nabla_{x}J^{1} + \nabla_{y}J^{2} \right) = 0 \to \int d^{2}x \frac{\partial}{\partial t}J^{0}=0 \quad \text{or} \quad \frac{\partial}{\partial t}\int d^{2}x J^{0}=0.
\end{equation*}
In the case of scale invariance, the Noether charge is  
\begin{equation}\label{generator D app.}
D = \int d^{2}xJ^{0} =2tH + \frac{i}{2}\int \vec{x}\left(\psi^{\ast}\vec{\nabla}\psi - \psi\vec{\nabla}\psi^{\ast} \right)d^{2}x.
\end{equation}
The second term in the r.h.s. of Eq.(\ref{generator D app.}) is time independent for static/stationary solutions (e.g. ansatz in (\ref{ansatz_background})). Additionally, the Hamiltonian of a system is also time-independent. Thus, the $\dot{D}$ for the stationary configurations is
\begin{equation}
    \frac{dD}{dt}=2H.
\end{equation}
From the other side, the conservation law is $\dot{D}=0$, which results in the condition $H=0$ for any stationary solution with arbitrary winding number $n$.
%\end{comment}

Now, we draw attention to a differential relation between energy $E$ and global $U(1)$ charge $Q$ of the Q-tube solutions
\begin{equation}
    \frac{dE}{d\omega} = \omega \frac{dQ}{d\omega}.
\end{equation}
One can explicitly check, that this relation is secured by equations of motion.

Finally, in the main text we have presented our results for numerical scanning of normalizable decay modes about a soliton solutions in a model (\ref{lagrangian}). As can be seen in Fig.\ref{modes}, the decay rate $\gamma_{dec.}$ of linear perturbations on the Q-tube background becomes significantly smaller with larger winding number $n$. A naive interpretation of this result comes from indicating that exponential clumping with factor $e^{\gamma_{dec.}t}$ may be suppressed due to the large intrinsic angular momentum of a Q-tube. By definition, intrinsic angular momentum
\begin{equation}
    J = \int T_{0\theta} d^{2}x,
\end{equation}
where $(0,\theta)$-component of the energy-momentum tensor is
\begin{equation*}
    T_{0\theta} = \partial_{t}\phi\partial_{\theta}\phi^{\ast} + \partial_{\theta}\phi\partial_{0}\phi^{\ast}.
\end{equation*}

Further calculations show that
\begin{equation}
    |J|=\biggl|\int \left[-\omega n f^{2}(r) \right]d^{2}x\biggr|=nQ.
\end{equation}
In the limit $\omega\to m$, large $n$ approximation from the previous appendix results in
\begin{equation}
    |J|=nQ_{max}\approx \frac{8n\pi m}{\lambda}\ln{\left(1+e^{2\sqrt{3}n}\right)}\to\frac{16\sqrt{3}\pi m}{\lambda}n^{2} \quad,\quad n\to \infty.
\end{equation}

\bibliography{biblio}
\end{document}